\newcommand{\tr}{\mathrm{tr}}  

\documentclass[
pre,twocolumn
,superscriptaddress
]{revtex4}
\usepackage{amssymb}
\usepackage{graphicx}
\usepackage{subfigure}
\usepackage{psfrag}

\psfrag{DD}{$\Delta$}
\psfrag{i=k}{$ i = k $}
\psfrag{j=l}{$ j = l $}
\psfrag{++}{$+$}
\psfrag{==}{$=$}

\psfrag{Delta}{$\Delta$}
\psfrag{cdots}{$\cdots$}
\psfrag{Quark propagators}{``Quark'' propagator $G_{ij}$}
\psfrag{Gluon propagators}
{``Gluon'' propagator $V^{1/2}_{ij}\Delta_{ij,kl}V^{1/2}_{kl}$}
\psfrag{(a)}{(a)}
\psfrag{(b)}{(b)}
\psfrag{(c)}{(c)} 
\psfrag{ii}{$i$}
\psfrag{jj}{$j$}
\psfrag{kk}{$k$}
\psfrag{ll}{$l$}
\psfrag{mm}{$m$}
\psfrag{nn}{$n$}
\psfrag{oo}{$o$}
\psfrag{pp}{$p$}
\psfrag{LL}{$L$}
\psfrag{Our diagrams}{Diagrams used}
\psfrag{paper}{ in this paper}
\psfrag{'t Hooft diagrams}{ \mbox{'t Hooft} diagrams}

\begin{document}

\title{A steepest descent calculation of RNA pseudoknots}

\author{M. Pillsbury}
\affiliation{Department of Physics, University of California, Santa
Barbara, CA 93106, USA}

\author{Henri Orland}
\affiliation{Service de Physique Th\'eorique, CEA Saclay, 91191
Gif-sur-Yvette Cedex, France} 
 
\author{A. Zee}
\affiliation{Department of Physics, University of California, Santa
Barbara, CA 93106, USA}
\affiliation{Institute for Theoretical Physics, University of
California, Santa Barbara, CA 93106, USA}

\begin{abstract}
  We enumerate possible topologies of pseudoknots in single-stranded
  RNA molecules. We use a steepest-descent approximation in the large
  $N$ matrix field theory, 
  and a Feynman diagram formalism to describe the resulting pseudoknot
  structure. 
\end{abstract}

\maketitle

An RNA molecule is a heteropolymer strand made up of four types of
nucleotides, uracil ($U$), adenine ($A$), guanine ($G$), and cytosine
($C$). The sequence of these nucleotides, or bases, makes up the
molecule's primary structure. Bases form hydrogen bonds with each
other to give the molecule a stable shape in three dimensions, with
$U$ bonding to $A$, and $C$ to $G$. Calculating the shape a given
primary structure will fold into is important in molecular biology.




We can associate $-U_{ij}$ with the energy of forming a hydrogen bond
between the $i$th and $j$th bases, and let $V_{ij}= \exp (U_{ij}/ T)$
where $T$ is the temperature. This is a minimalist model: we make no
attempt to account for loop penalties or stacking interactions. There
is some rigidity in the chain of nucleotides, as well as steric
constraints, which prevent hydrogen bonding between nucleotides that
are within four bases of each other, so we let $ V_{i,i+k} = 0$ if $k
< 4$.  The partition function associated with this bonding is given by

\begin{eqnarray}
  Z_{L,1}
  & = & 1+ \sum_{i<j} V_{ij} + \sum_{i<j<k<l} V_{ij}
  V_{kl} + \sum_{i<j<k<l} V_{ik} V_{jl}\nonumber\\
  & + & \sum_{i_1<i_2< \dots <i_n}V_{i_1i_2}V_{i_3i_4}\dots V_{i_ni_{n+1}}
  + \ldots \label{eq:chp}
\end{eqnarray}

\noindent
Evidently, $Z_{L,1}$ is the combinatorial heart of the RNA folding
problem \cite{oz}. While $Z_{L,1}$ appears very simple at
first glance, it contains a term for every possible configuration of
bonds on the chain. Finding the folded state could involve searching
through $ \sim L!$ terms, which is a daunting task for even the
shortest RNAs.

Fortunately, in RNA, there is a hierarchical separation between
primary, secondary and tertiary structures that reduces the number of
configurations that must be considered.  One can find the secondary
structure by drawing the chain of nucleotides around the circumference
of a circle, with the first nucleotide next to the last, and finding a
bond structure that minimizes the free energy with the constraints
that all bonds are drawn as arcs within the circle, and no bonds
cross.  Another representation is to draw the bond structures as
systems of parallel arches which do not cross.
This planar configuration (in the sense used in \cite{nj,
  coleman}, though other usages are common in the RNA folding
literature) is made up of the secondary structure's characteristic
loops and bulges.  
Bonds between distinct parts of the secondary
structure are called pseudoknots, and are typically considered part of
the molecule's tertiary structure. For instance, the contributions
from the third sum in \ref{eq:chp} come from pseudoknot
configurations.  The formation of the tertiary structure is believed
not to alter the more stable secondary structure \cite{higgs, tinoco}.

Secondary and tertiary structures are usually stable at biological
temperatures, which are typically well below the RNA molecule's
melting point.  This makes certain very efficient algorithms for
determining RNA secondary structure at zero temperature possible and
useful. These ``dynamic programming'' methods involve recursively
calculating $Z_{L,1}$ and then backtracking to find the dominant terms,
and thus determine which bonds are present in the folded RNA.  There
are also dynamic programming techniques that try to account for
pseudoknots, but they are necessarily slower\cite{nj,smith,rivas}.

The distinction between secondary and pseudoknot structure has a
topological flavor. One powerful tool for dealing with topological
considerations is the large $N$ expansion used in matrix field
theories.  Originally proposed by \mbox{'t Hooft} to represent quantum
chromodynamics with $N$ colors, it predicts that non-planar Feynman
diagrams have amplitudes proportional to negative powers of $N$, and
are thus suppressed when $N$ is large\cite{thooft, coleman}. Two of
the authors applied a similar technique to the problem of RNA folding,
leading to the same sort of suppression of non-planar configurations;
we summarize the results below, and refer the reader to \cite{oz} for
details.

One can perform a series of manipulations to find that a chain of $L$
bases has 

\begin{equation} \label{eq:part}
Z_{L,1} = \frac{1}{C} \int dA\, e^{-\frac{N}{2} \left( \tr A^2 +2 \tr
    \log M(A) \right)} M^{-1} (A)_{L+1,1} 
\end{equation}

\noindent 
where the integral is taken over all Hermitian $(L+1) \times (L+1)$
matrices $A$. $C$ is an unimportant normalization constant and $M$ is
a matrix function of $A$ given by

\begin{equation}
M_{ij} =\delta_{ij} - \delta_{i,j+1} +i \sqrt{V_{i-1,j}}\, A_{i-1,j}
\end{equation}

\noindent 
Here, $N$ is a used to keep track of topology. as mentioned above.
Thus we can expand in powers of $1/N$ and evaluate the integral by
steepest descent.  We need to find the stationary point of the
``action''

\begin{equation}
S(A) \equiv \frac{1}{2} \tr\, A^2 - \tr\, \log M(A)
\end{equation}
 
\noindent which requires solving 
$ \frac{\delta S(A)}{\delta A} = 0$. 
This occurs at the point $\tilde{A}$, which is defined by

\begin{equation} \label{eq:stpt}
\tilde{A}_{lk} = i \sqrt{V_{lk}}\, (M^{-1})_{l,k+1}
\end{equation}

We define a new matrix in terms of $M^{-1}$ at the
stationary point,

\begin{equation} \label{eq:stpt1}
G_{ij} = (M^{-1})_{i+1,j}
\end{equation}

\noindent and use the trivial identity 
$ \sum_j M_{ij} (M^{-1})_{jk} = \delta_{ik}$ to derive the Hartree
equation

\begin{eqnarray}
 G_{i+1,k}  & = & \delta_{i+2,k} + G_{ik} \nonumber \\
            & + & \sum_j V_{i+1,j} G_{i,j+1} G_{j-1,k} \label{eq:hf}
\end{eqnarray}

\noindent This equation is recursive, and we need to impose the
boundary condition that $G_{i,i+l} = 0 $ for $ l \ge 2 $ to solve it.
Then, $G_{ij}$ is the partition function of the helical secondary
structure of a chain that starts with the $j$th base and ends with the
$i$th base. This form is precisely that used in existing dynamic
programming algorithms\cite{higgs,nj,smith}. Since it carries two
indices, $G_{ij}$ is analogous to the quark propagator in large $N$
QCD, which carries two indices for color. The recursion relation
ensures that it is a ``dressed'' propagator.

We can then introduce the fluctuation $x_{ij}$, defined by $A_{ij} =
\tilde{A}_{ij} + x_{ij}/\sqrt{N}$, and expand $\tr \log (M^{-1}(A))$
and $M^{-1}(A)$ as power series in $x$.  Then we collect powers of
$N^{-1/2}$ to find corrections to the steepest descent approximation
of $Z_{L,1}$. We are left with Gaussian integrals in $x_{ij}$ that can
be evaluated by applying Wick's theorem, with 
contractions given by the inverse of the quadratic
form in the exponential. This inverse is a propagator which satisfies
the Bethe-Salpeter equation

\begin{eqnarray} \label{eq:bs}
\Delta_{kl,mn} & = & \delta_{km} \delta_{nl} \nonumber \\
        & + & \sum_{ij} V_{kl}^{1/2} V_{ij}^{1/2} G_{k-1,i+1}
        G_{j-1,l+1} \Delta_{ij,mn}
\end{eqnarray}

\noindent 
While the Hartree equation gave the partition function for a single
contiguous chain of RNA interacting with itself, the Bethe-Salpeter
relation gives the contribution from two separated segments. 

\begin{figure}[htp] 
   \centering
   \subfigure[Bethe-Salpeter relation]{
      \includegraphics[width=0.4\textwidth]{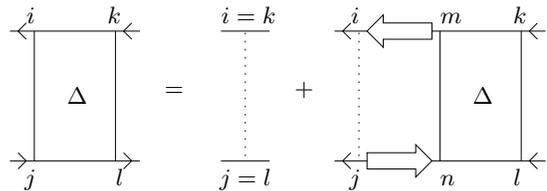}
      \label{fig:bs:eqn}}
   \subfigure[Sample structure contributing to $ \Delta $.]{
      \includegraphics[width=0.25\textwidth]{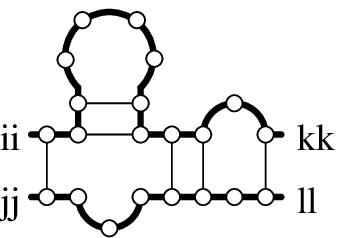}
      \label{fig:bs:samp}}
   \caption{The $\Delta$ propagator.}
   \label{fig:bs}
\end{figure}

Physically, $\Delta$ represents the resummation of all ladder diagrams
between anti-parallel segments, where each segment is itself dressed
by secondary structure elements. Equation (\ref{eq:bs}) can be
represented pictorially, as in fig.~\ref{fig:bs:eqn}.
There are four indices on
$\Delta_{ij,kl}$, indicating where the segments begin and end, so we
call it a ``gluon propagator'', in analogy to gluon propagators in
QCD, which carry four color indices. A typical structure contributing to
$\Delta$ is shown in fig.~\ref{fig:bs:samp}.
Each single line propagator is dressed by any system of arches.  There
can be any number of parallel interactions between the two strands.
The only constraint is that no interaction lines should cross.  Note
that a system of arches along one line (RNA strand) typically
represents a piece of helix on this strand, whereas a system of
parallel interactions between the two lines can represent a helical
fragment between the two strands.

\begin{figure}[b!] 
   \centering
    \includegraphics[width=0.4\textwidth]{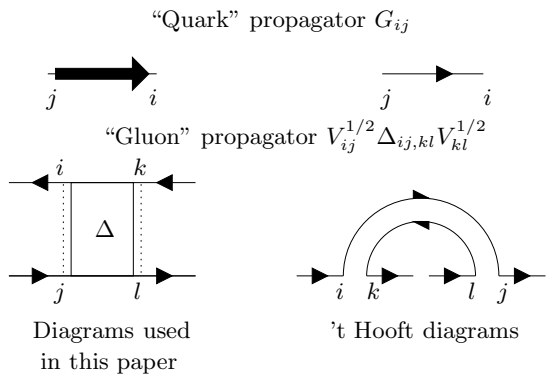}
    \caption{Propagators.}
    \label{fig:prop}
\end{figure}

There are two ways of drawing Feynman diagrams for these propagators.
The first was introduced in \cite{oz}, and is useful for visualizing
the RNA's topology.  The second is the double-line formalism of
\mbox{'t Hooft}, which makes it very easy to find a graph's order in
$1/N$, by assigning appropriate powers powers of $N$ to loops, edges
and vertices\cite{coleman,thooft}.  It follows from (\ref{eq:bs}) that
the $\Delta$ propagator contains powers of $V^{1/2}_{ij}$, but the
partition function (\ref{eq:chp}) contains only whole powers of
$V_{ij}$. Thus, all $\Delta$'s in the expansion appear with factors of
$V^{1/2}_{ij}$, as $ V^{1/2}_{ij} \Delta_{ij,kl} V^{1/2}_{kl} $.  This
is reflected in the diagrams in fig. \ref{fig:prop}. We can then
expand $Z_{L,1}$ to order $N^{-2}$, getting the secondary structure as
well as the tertiary correction to it. Then

\begin{widetext}

\begin{eqnarray} \label{eq:sec}  
 Z_{L,1} & = & G_{L,1} \\
        & + & \frac{1}{N^2} \left\langle \left[ 
        \left( B_4 -\frac{1}{4} B_2T_4 
        -\frac{1}{3} B_3 T_3 -\frac{1}{5} B_1 T_5
        +       
        \frac{1}{12} B_1 T_3 T_4 + \frac{1}{18} B_2 T_3^2 -
        \frac{1}{162} B_1 T_3^3 \right) M^{-1}  
        \right]_{L+1,1} \right \rangle \label{eq:ter}
\end{eqnarray}

\end{widetext}

\noindent 
where we use the value of $M^{-1}$ at the stationary point from
(\ref{eq:stpt},\ref{eq:stpt1}). We have also introduced some
convenient shorthand for matrices and traces that contain powers of
$x$,

\begin{eqnarray*}
  c_{ij} & = &
      \sqrt{V_{i-1,j}} x_{i-1,j} \\
  D_{mn} & = & \sum_{m'} (M^{-1})_{mm'} c_{m'n} \\
  (B_p)_{kl} & = & (D^p)_{kl} \\ 
  T_p & = & \tr\, B_p
\end{eqnarray*}




\noindent 
The angle brackets in (\ref{eq:ter}) mean the included terms should be
integrated over $x_{ij}$ with the Gaussian weight $\exp [-(\tr\,
x^2+\tr(M^{-1} c)^2)/2]$. These integrals are simple in principle, as
the $x_{ij}$'s can be contracted with the Bethe-Salpeter propagator
(\ref{eq:bs}). Each power of $x$ introduces a vertex for gluon lines.

The multiplication implicit in the definition of $B_p$ is matrix
multiplication, so many indices must be summed over when evaluating
the terms in (\ref{eq:ter}). For instance, evaluating one of the
contractions of $ \langle B_4 M^{-1}\rangle $ produces the sum, 
\begin{eqnarray}
  \left\langle \left(B_4 M^{-1}\right)_{L+1,1} \right\rangle &=& 
  \sum_
  {
    \begin{array}{c}
      \scriptstyle{i,j,k,l,}\\
      \scriptstyle{m,n,o,p}
    \end{array}
    }
  G_{L,i+1} G_{j,k+1} \nonumber\\
  & \times &  G_{l, m+1} G_{n, o+1} G_{p,1} \nonumber\\ 
  & \times & V^{1/2}_{i,n+1} \Delta_{i,n+1,j+1,m} V^{1/2}_{j+1,m} 
  \nonumber\\
  & \times &   V^{1/2}_{k,p+1} \Delta_{k,p+1,l+1,o} V^{1/2}_{l+1,0}
 \label{eq:b4}
\end{eqnarray}
\noindent 

\begin{figure}[hbt]
  \centering
  \resizebox{0.2\textwidth}{!}{\includegraphics{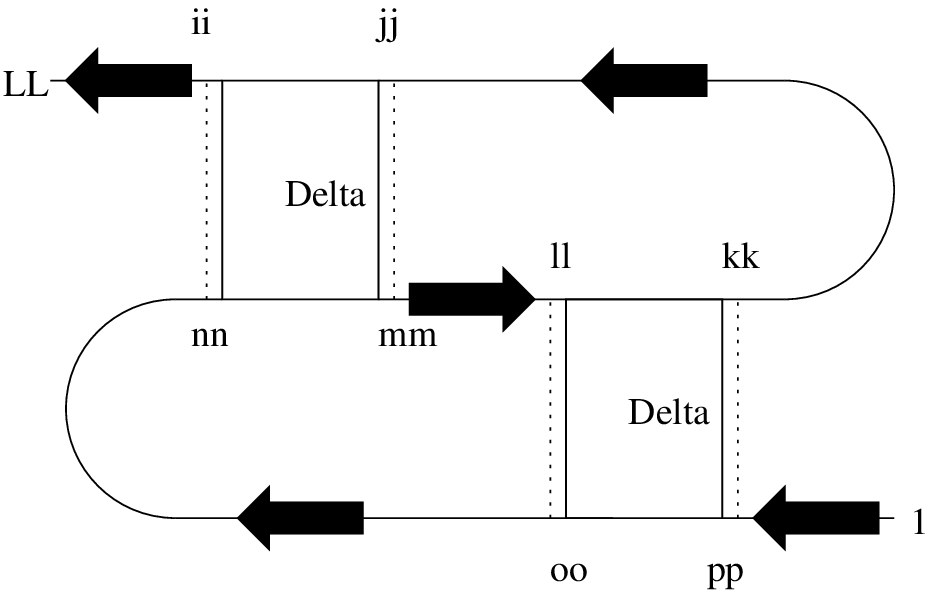}}
  \caption{ Diagram for $ 
    \left\langle \left[B_4 M^{-1}\right]_{L+1,1} \right\rangle $}
  \label{fig:b4}
\end{figure}

\noindent
Looking at the diagram associated in the contraction in fig.
\ref{fig:b4}, and using the condition that $G_{a,a+b} = 0$ for $b \ge 2$,
we deduce the proper constraint for the indices, $ L \ge i > j \ge k >
l \ge m > n \ge o > p \ge 0 $ .

\begin{figure}[htp] 
   \centering
   \subfigure[$B_m$]{
      \scalebox{0.35}{\includegraphics{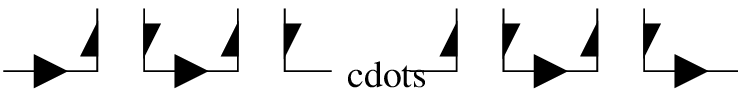}}
      \label{fig:tb:a}}
   \subfigure[$T_n$]{
      \scalebox{0.35}{\includegraphics{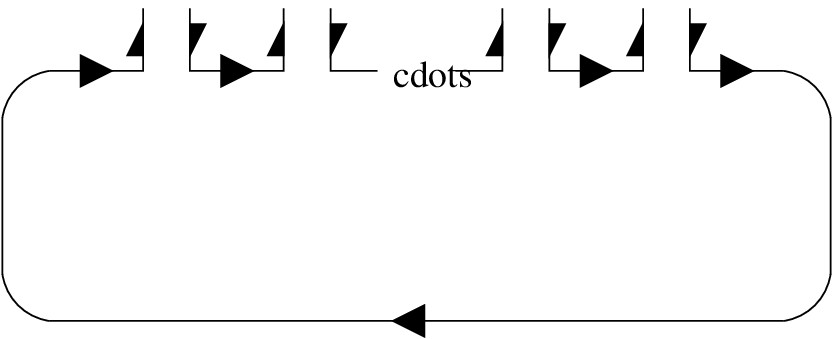}}
      \label{fig:tb:b}}
   \caption{Matrix products}
   \label{fig:tb}
\end{figure}

The $B_m$ and $T_n$ terms have simple \mbox{'t Hooft} diagrams, as
shown in fig. \ref{fig:tb}. The ellipses in the diagram represent the
string of $m$ or $n$ gluon vertices associated with those terms.  The
graph for $T_n$ closes on itself, reflecting the trace's cyclic
symmetry.

These diagrams make it simple to pick out the Wick contractions that
actually contribute to $Z_{L,1}$. One can draw Feynman diagrams for the
contractions of the 7 terms in (\ref{eq:ter}), and find that 25 of
them are distinct (many contractions are equivalent under the cyclic
symmetry of the traces $T_n$). However, most of these vanish, as they
contain closed $G$ loops.  Diagrams involving closed loops will depend
on a factor of $G_{i,i+l}$ for $l > 2$, and therefore vanish. This can
also be understood in terms of the diagrams from \cite{oz}, where
$G$'s represent segments of RNA, and $\Delta$'s represent interactions
between two segments. A closed $G$ loop with both ends connected to
the same side of a $\Delta$ propagator describes a closed loop of RNA
interacting with the main strand. We have specifically excluded this
possibility from our definition of $Z_{L,1}$, so such configurations
must vanish. This is the reason why there is no graph of order $1/N$
in (\ref{eq:sec}).

\begin{figure}
   \includegraphics[width=0.35\textwidth]{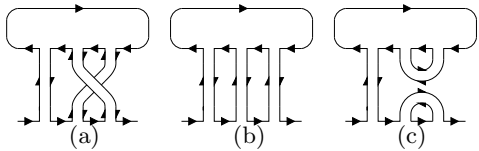}
   \caption{Contractions for $ \langle B_3 T_3 M^{-1} \rangle $}
   \label{fig:abcbca}
\end{figure}
 
As an example, consider $ \langle B_3 T_3 M^{-1} \rangle $, which can
be contracted in the three distinct ways shown in figs.
\ref{fig:abcbca}(a), (b) and (c). Each of these occurs with a symmetry
factor of 3, since an $x_{ij}$ from the $B_3$ can be contracted with
any of the (cyclically equivalent) $x_{ml}$'s in $T_3$. Only the
diagram in fig. \ref{fig:abcbca}(a) can be traced with an unbroken
line---the other diagrams contain closed loops. Thus, only one of the
three sorts of contractions contributes to the partition function.

\begin{figure*}
   \centering
   \subfigure[$ B_4 M^{-1} $]{
      \scalebox{0.26}{
      \includegraphics{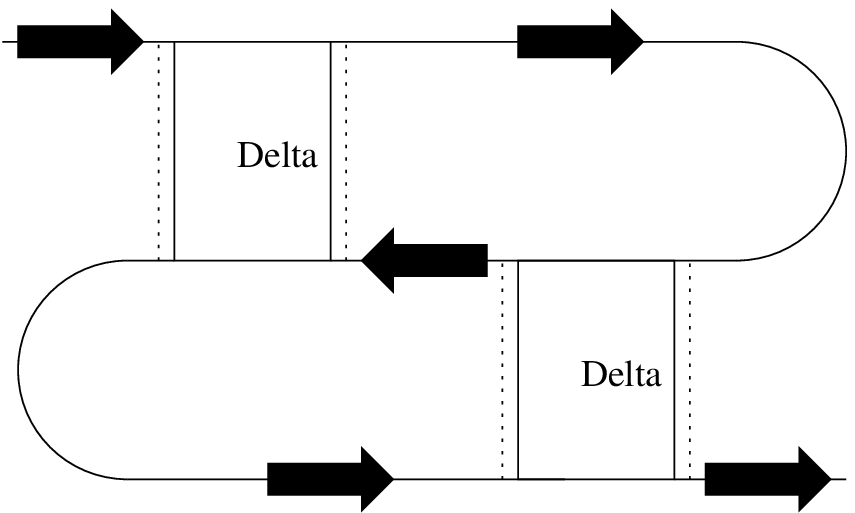}
      \label{fig:folds:a}}}
   \subfigure[$ B_2 T_4 M^{-1} $]{
      \scalebox{0.26}{
      \includegraphics{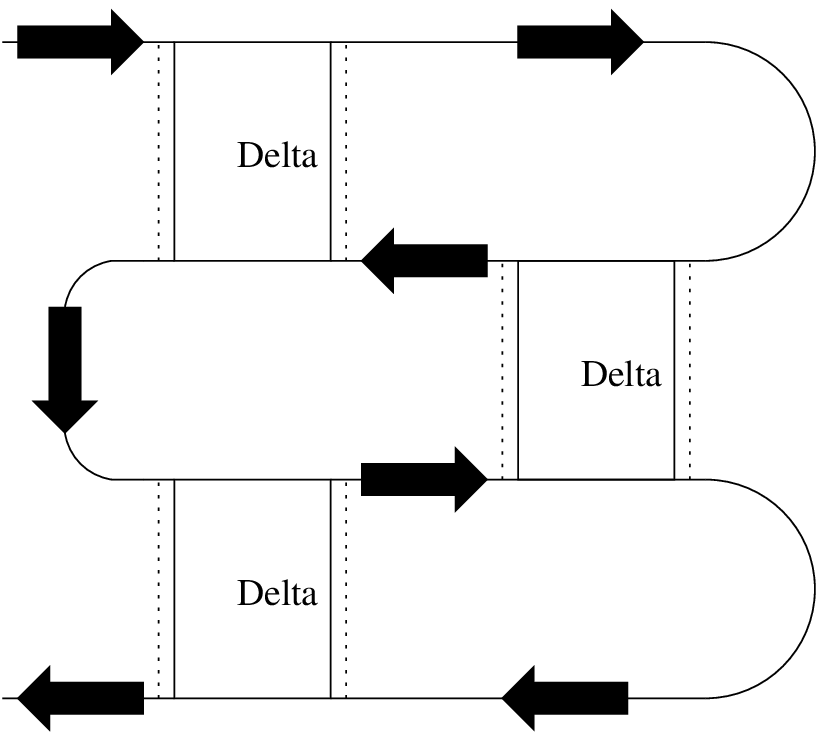}
      \label{fig:folds:b}}}
   \subfigure[$ B_3 T_3 M^{-1} $]{
      \scalebox{0.26}{
      \includegraphics{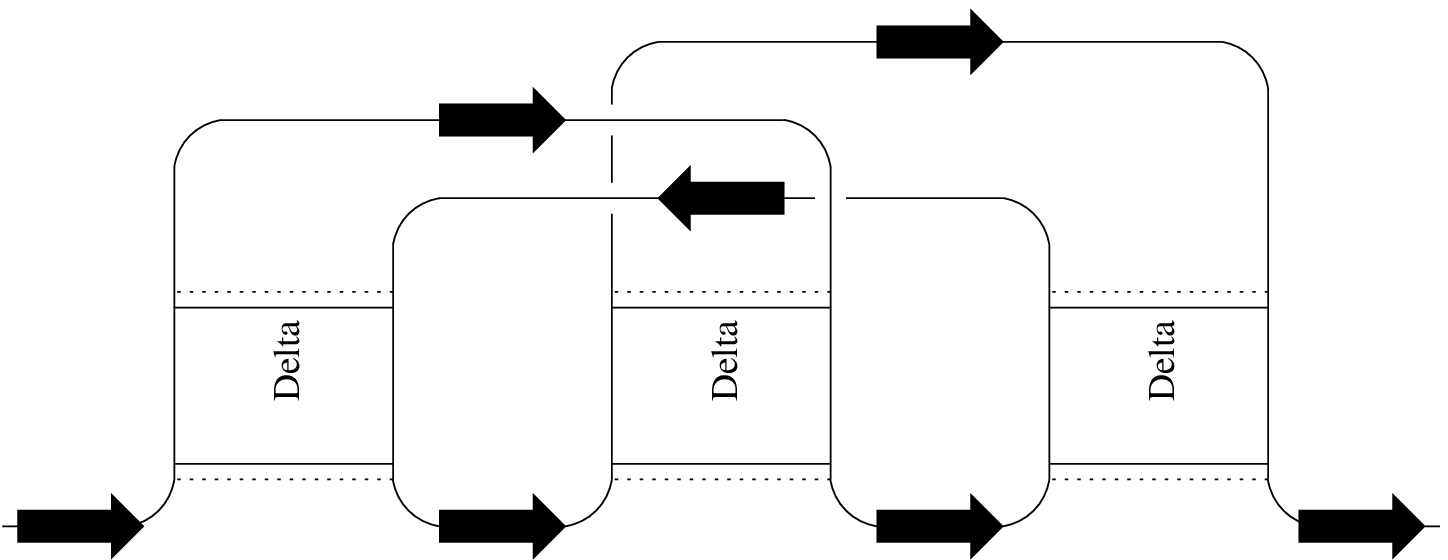}
      \label{fig:folds:c}}}
   \subfigure[$ B_1 T_5 M^{-1} $]{
      \scalebox{0.26}{
      \includegraphics{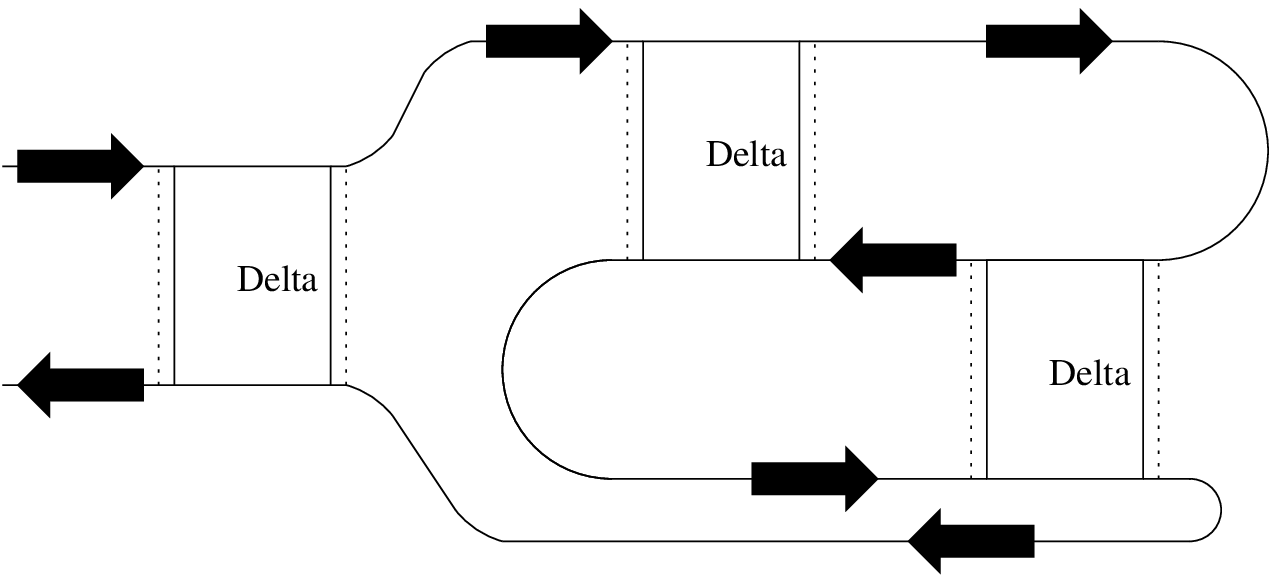}
      \label{fig:folds:d}}}
   \subfigure[$ B_1 T_3 T_4 M^{-1} $]{
      \scalebox{0.26}{
      \includegraphics{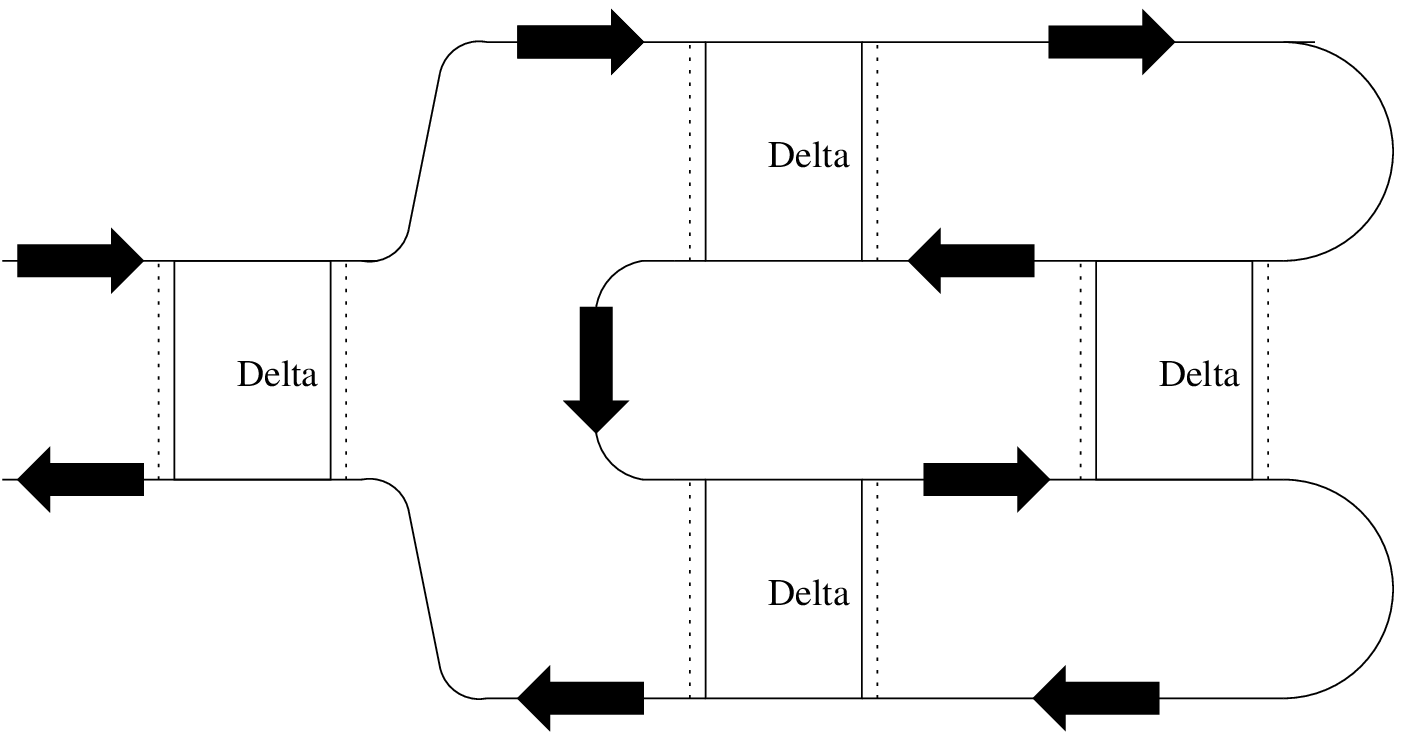}
      \label{fig:folds:e}}}
   \subfigure[$ B_1 T_3 T_4 M^{-1} $]{
      \scalebox{0.26}{
      \includegraphics{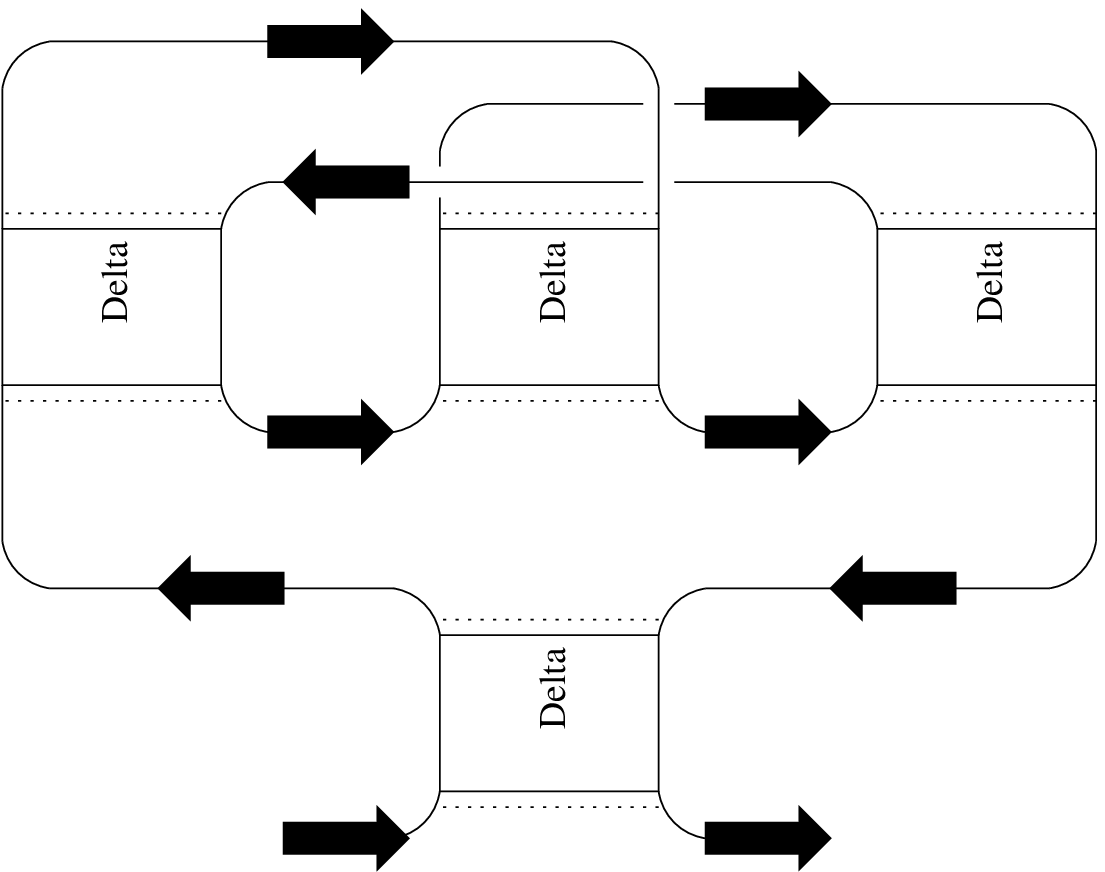}
      \label{fig:folds:f}}}
   \subfigure[$ B_2 T_3^2 M^{-1} $]{
      \scalebox{0.26}{
      \includegraphics{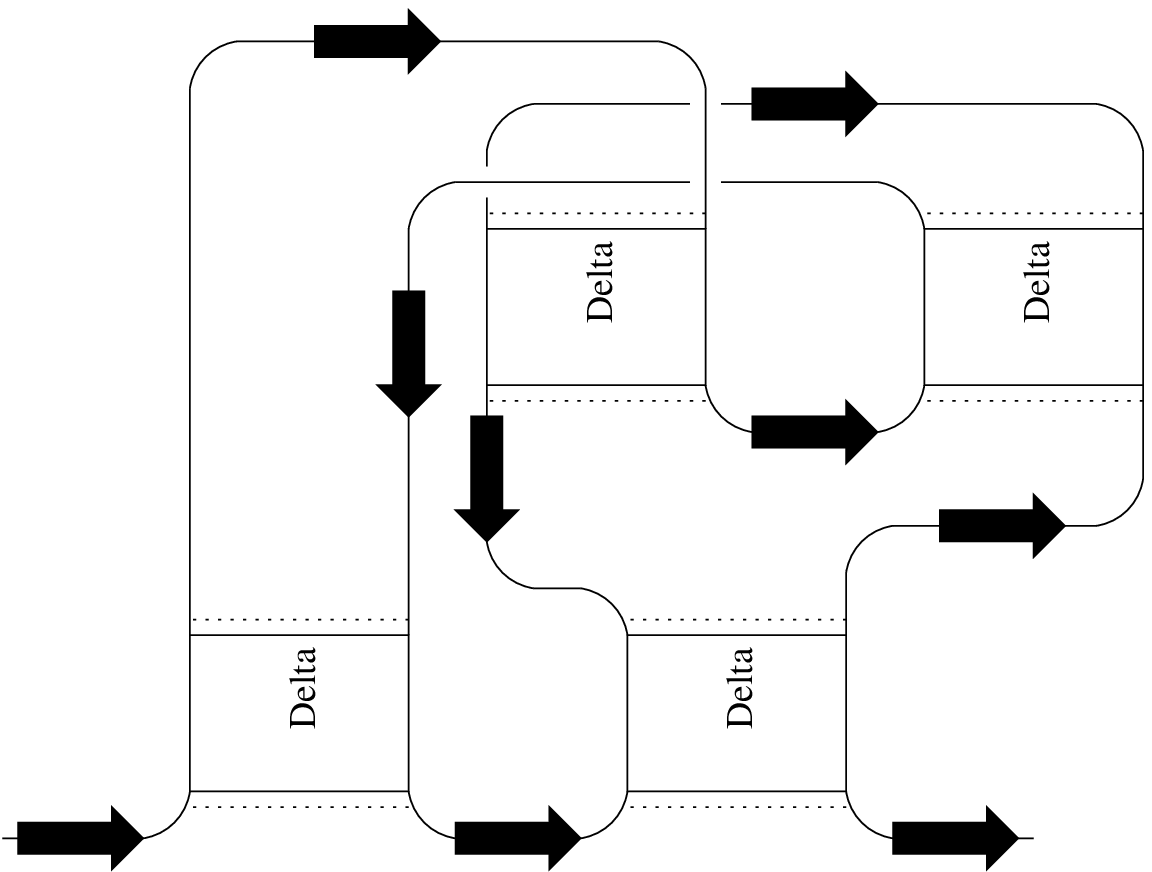}
      \label{fig:folds:g}}}
   \subfigure[$ B_1 T_3^3 M^{-1} $]{
      \scalebox{0.26}{
      \includegraphics{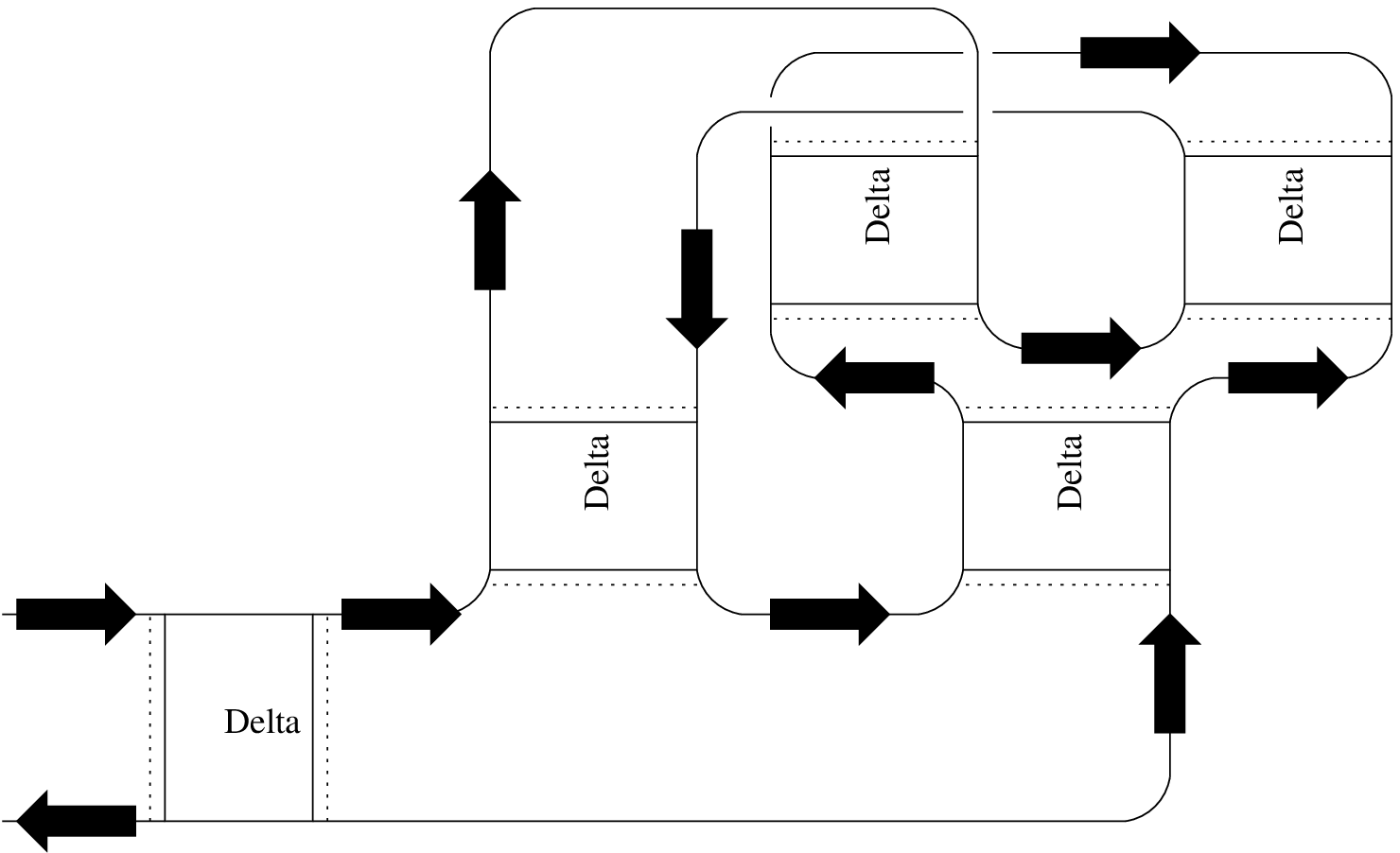}
      \label{fig:folds:h}}}
   \caption{Non-vanishing contractions}
   \label{fig:folds}
\end{figure*}

When all the contractions have been carried out, there remain 8
non-vanishing graphs, which are shown in fig. \ref{fig:folds}. The
contractions associated with each diagram are

\begin{center}
\begin{tabular}{|c|c|l|}
\hline
\textbf{Figure} & \textbf{Contraction} & \textbf{Pseudoknot} \\
\hline
(a) & $ B_4 M^{-1} $ & ABAB \\
\hline
(b) & $ B_2 T_4 M^{-1} $ & ABACBC \\ 
\hline
(c) & $ B_3 T_3 M^{-1} $ & ABCABC \\ 
\hline
(d) & $ B_1 T_5 M^{-1} $ & ABCBCA \\ 
\hline
(e) & $ B_1 T_3 T_4 M^{-1} $ & ABCBDCDA \\
\hline
(f) & $ B_1 T_3 T_4 M^{-1} $ & ABCDBCDA \\
\hline
(g) & $ B_2 T_3^2 M^{-1} $ & ABCADBCD \\
\hline
(h) & $ B_1 T_3^3 M^{-1} $ & ABCDBECDEA \\
\hline
\end{tabular}
\end{center}

\noindent The alphabetic notation, common in the biochemical
literature, shows the order in which sites pair with each other. For
example, ``ABAB'' indicates that the first and third vertices (both
denoted by ``A'') are paired, and that the vertex between them is
linked to the fourth vertex (both denoted by ``B'').

Since the pseudoknots we consider contribute to order $1/N^{2}$, only
one pseudoknot may be present at a time. This problem can be solved by
noting that all the pseudoknot diagrams are one particle irreducible
(1PI, i.e. they cannot be disconnected by opening a single quark
line), and can thus be re-summed by a Dyson equation.  Define
$\Sigma_{mn}$ as the sum of all the amputated pseudoknot diagrams
defined above (i.e. the sum of all $O(N^{-2})$ 1PI diagrams with their
external $G$ propagators removed). Then the partition function
$Z_{mn}$ satisfies the usual Dyson equation:

\begin{equation}
Z_{mn}  =  G_{mn} + \sum _{m<k<l<n} Z_{mk}\Sigma_{kl}
       G_{ln}\label{eq:dyson}
\end{equation}

\noindent
Once the 8 diagrams for $\Sigma $ have been calculated, the full
partition function (with any number of pseudoknots) can be calculated
using the above recursion relations. Present knot-prediction
algorithms use dynamic programming allow knots which have bonds drawn
inside and outside of the disc, as long as they are no
crossings\cite{rivas}. This excludes certain topologies our algorithm
provides for, like ABCABC pseudoknots. On the other hand, these
algorithms do provide for some topologies that we've excluded as $
\mathcal{O}(N^{-4}) $. 

The method presented allows us to calculate the partition function in
$ O(L^6) $ time, so it can be used for folding, by backtracking to
pick out the largest term in the partition function. The strategy for
doing so is the following: i) solve for the Hartree partition function
(\ref{eq:hf}), ii) solve the Bethe-Salpeter recursion equation
(\ref{eq:bs}) to get $\Delta _{kl,mn}$, iii) calculate the eight
amputated diagrams of fig. 5 making up the 1PI function $\Sigma_{mn}$,
iv) solve the Dyson equation (\ref{eq:dyson}) by recursion to obtain
the full partition function with any number of pseudoknots, v) and
then backtrack to find the largest term in this partition function.

Some numerical calculations are under way and we hope to present those
results in a future paper, along with an explicit calculation for the
order $N^{-2}$ folding of a short ($L\simeq 10$) RNA.

\end{document}